\documentclass[prd,tightenlines,floatfix,preprint,nofootinbib,eqsecnum,12pt]{revtex4-1}

\usepackage{amsmath}

\def\1p{{(1p)}}

\def\xtypO{\xi^{\rm typO}}
\def\xtypD{\xi^{\rm typD}}
\def\pe{p_E}

\begin{document}

\title{Science in a Very Large Universe}

\author{Mark Srednicki}
\email{mark@physics.ucsb.edu}

\author{James Hartle}
\email{hartle@physics.ucsb.edu}

\affiliation{Department of Physics, University of California, Santa Barbara, CA 93106-9530}

\begin{abstract}
As observers of the universe we are quantum physical systems within it. If the universe is very large in space and/or time, the probability becomes significant that the data on which we base predictions is replicated at other locations in spacetime. The physical conditions at these locations that are not specified by the data may differ. Predictions of our future observations therefore require an assumed probability distribution (the {\it xerographic distribution}) for our location among the possible ones. It is the combination of basic theory plus the xerographic distribution that can be predictive and testable by further observations.
\end{abstract}


\maketitle

\section{Introduction}

Theories of our universe are tested using the data that we acquire. When calculating predictions, we customarily make an implicit
assumption that our data $D_0$ occur at a unique location in spacetime. However, there is a quantum probability for these data to exist in any spacetime volume.  This probability is extremely small in the observable part of the universe. However, in the large (or infinite) universes considered in contemporary cosmology, the following predictions often hold:
\begin{enumerate} 
\item The probability is near unity that our data $D_0$ exist somewhere.
\item The probability is near unity that our data $D_0$ is exactly replicated elsewhere many times. An assumption that we are unique is then false. 
\end{enumerate}
This paper is concerned with the implications of these two statements for science in a very large universe.  Some implications of the first were discussed in \cite{HS07}. We shall return to these below, but we first focus on the implications of the second.


\section{Third-Person and First-Person Prediction}

It is useful to distinguish between first-person and third-person predictions in cosmology. Third-person predictions are made through the probabilities for alternative features of the universe that  it may exhibit. Examples are the probabilities that the universe is homogeneous and isotropic, that it had a certain number of inflationary $e$-folds in the past,  that it will end in a big crunch, or that it exhibits a certain number of instances of our data $D_0$. Familiar quantum theories of the universe make such third-person predictions by specifying a quantum state and a prescription for dynamics (see e.g. \cite{HHH}); such theories are denoted by $T$. But to use and test theories we need predictions of what {\it we} will measure or predict. These are first-person predictions. Defining these is not trivial if there is more than one instance of our data $D_0$.  We now discuss how to do it. 

As observers of the universe we are a subsystem within it that we call the human scientific IGUS\footnote{IGUS is an acronym for Information Gathering and Utilizing System.} (HSI).  Terms like `we', `us', `our' refer to this specific subsystem.  The HSI can be described at various levels of coarse-graining. Here it is assumed that a description is fixed at a classical level\footnote{More precisely we assume that there is a description in terms of the quasiclassical variables that describe the quasiclassical realm of everyday experience in terms of a decoherent set of coarse-grained alternative histories defined in terms of these variables; see e.g.~\cite{Har09} for more detail.}. This description can be divided into two parts:  First the data $D_0$ that the HSI has: every scrap of information that the HSI possesses about the physical universe, including every record of every experiment, every astronomical observation of distant galaxies, every available description of every leaf, etc., and necessarily every piece of information about the HSI itself, its members, and its history. Second, there are the quantities  not included in the data but which are necessary for a complete physical description of the HSI.  Location in the universe is one example; physical circumstances that have not yet been measured is another.  (The HSI is
assumed to exist at a unique location in the universe but its data $D_0$ may be duplicated at many other locations in a very large universe.) 

First-person predictions are through the probabilities of what our specific instance of an IGUS with data $D_0$ will observe or measure.  But theories $T$ of a quantum state and dynamics do not make such predictions directly if the chance is significant that our data $D_0$ are replicated at different places in the universe. Third-person probabilities computed from $T$ make no reference to `us' and contain no information about which of several instances of $D_0$ is `us'. A further assumption is therefore needed to connect the third-person probabilities of theory with the first-person probabilities for our observations. We call this assumption the {\it xerographic distribution} and describe it in the next section.

\section{The Xerographic Distribution}

Consider a universe that contains $N$ copies of $D_0$ at different locations in spacetime $x_A$, $A=1,\ldots,N$. We are one of these copies, but we have no information as to which, since our data is replicated at each of these locations.  Therefore, to predict what we will observe in our future, we must choose a distribution that gives the probability $\xi_A$ that we (the HSI) are the copy located at $x_A$.  The probabilities $\xi_A$ constitute the {\it xerographic distribution.} Location is used here as an illustration; the xerographic distribution could also refer to any other aspect of our physical situation that is not specified by the data $D_0$.

As mentioned above, the xerographic distribution is {\it not\/} fixed by usual candidates for the theory $T$, such as an initial quantum state and a prescription for its evolution. The xerographic distribution is a {\it further assumption\/} that {\it must\/} be made (explicitly or implicitly) in order to make first-person predictions. What is tested by observation is not just the theory $T$, but rather the combination $(T,\xi)$; such a combination of a fundamental theory (including the initial state) $T$ and the xerographic distribution $\xi$ will be called a {\it theoretical framework} or {\it framework} for short\footnote{A good case can be made for calling the combination $(T,\xi)$ the `theory', as it is the collection of assumptions from which testable predictions are made.  But to do so would risk confusion with the usual notion of a theory as consisting of fundamental laws for dynamics and an initial state. Furthermore the two parts of the combination $(T,\xi)$ are of different character. The theory $T$ supplies third-person probabilities for features of the whole universe. The xerographic distribution refers to assumptions about a particular subsystem: the HSI. For these reasons, in this paper `theory' is used in the usual way, and the combination $(T,\xi)$ is called a `theoretical framework'.}.

A natural assumption is that we are typical of the instances of $D_0$; this implies a uniform xerographic distribution $\xi_A=1/N$. But it is also possible to assume that we are {\it a}typical. We argued in \cite{HS07} that typicality was no more motivated by observation than atypicality.

More generally, the xerographic distribution can be used to express typicality assumptions that involve data sets other than $D_0$. For instance, as we have defined it, $D_0$ includes the results of our observations as well as the conditions for them.  But for some purposes it may be useful  to assume that our results are typical of all instances of similar observational situations. Or it may be useful to assume that observations are typical of those made by any IGUS in the galaxy, or by any IGUS in the universe. Each of these typicality notions corresponds to a xerographic distribution that is uniform on the class involved.  In the following we will show that different assumptions about $\xi$ can be  testable, and use this to  address the issues raised by the possibility of `Boltzmann brains'.

It is important to note that the use of a xerographic distribution does not constitute a modification of the laws of the quantum mechanics of closed systems. The usual laws, the usual rules for implementing them (including Born's rule), and the usual interpretation apply to third-person probabilities. First-person probabilities are the new feature. These are made necessary because we, like other IGUSes, are quantum subsystems of the universe with a non-zero probability of being replicated exactly elsewhere. Quantum theory must be augmented by a prescription to calculate the first-person probabilities. 
There is nothing mysterious or even especially quantum about this prescription because it operates at the classical level\footnote{Indeed it would be equally necessary if the theory $T$ were classical \cite{SH10}.}. Conversely we cannot expect to derive the xerographic distribution from the rules of quantum theory for third-person probabilities.

\section{Comparing Theories of Large Universes}

Many different criteria can be used by physicists to discriminate between  competing  theoretical frameworks on the basis of the available data $D_0$. Frameworks are favored  that are testable, simple, beautiful, precisely formulable mathematically, economical in their assumptions, comprehensive, unifying, explanatory, accessible to existing intuition, etc.  Most importantly,  scientists  favor frameworks that are successful in predicting new data beyond what they have at the moment. That is, they favor frameworks that are {\it predictive.} We will discuss this criterion more fully in Section \ref{pred}. In this section we discuss the simple idea that theoretical frameworks can be distinguished by the probabilities (likelihoods) that they give for the data $D_0$.

The process of distinguishing between frameworks by likelihoods is formalized in the widely familiar Bayesian schema for testing theories. While seldom applied in practice,  this schema is useful to state the assumptions made in the process clearly and quantitively. The process involves computing posterior probabilities for   frameworks  from the likelihoods for the data $D_0$ and a set of  prior probabilities for the competing frameworks\footnote{A brief review consistent  with present notation is given in \cite{HS07}.}. 

For a very large universe there are different possibilities for a Bayesian comparison of theoretical frameworks allowed by the distinction between first-person and third-person probabilities. We can seek to distinguish between theories $T$ on the basis of their third-person likelihoods for the data $D_0$ independent of any typicality assumption represented by a xerographic distribution. Or, we can seek to distinguish between different  frameworks $(T,\xi)$ on the basis of first-person probabilities  that follow from an assumption of a xerographic distribution. The question of which to use is moot in a small universe where the data $D_0$ are unique to a good approximation and first-person and third-person probabilities agree. However, the question becomes 
important in a large universe where the data $D_0$ may be replicated many times. 

First, consider applying the Bayes procedure to distinguish between theories just on the basis of their third-person predictions for $D_0$. 
As we emphasized in \cite{HS07}, all we know for certain from our data $D_0$ is that the universe exhibits at least one instance of it. We do not know how many times it is replicated or how frequently it occurs. A theory $T$ is consistent with our data if the probability for at least one instance of our data is nonzero. But, as has been stressed by many, in a large universe the likelihood that at least one instance of our data exists somewhere approaches unity for any theory that is consistent with our data. The third-person Bayes procedure is therefore not effective for discriminating between theories in a very large universe. 

We therefore consider applying the Bayes schema to  frameworks $(T,\xi)$. This involves the following elements: First, prior probabilities $P(T,\xi)$ must be chosen for the different frameworks. Next, the first-person likelihoods $P^\1p(D_0|T,\xi)$ must be computed. Finally, the first-person posterior probabilities are given by 
\begin{equation}
\label{bayesformula}
P^\1p(T, \xi |D_0) = \frac{P^\1p(D_0|T,\xi)P(T, \xi)}{\sum_{(T,\xi)}P^\1p(D_0|T,\xi)P(T,\xi)}
\end{equation}
The larger these are,  the more favored are the corresponding framework. 

As already noted, the theory $T$ and the xerographic distribution $\xi$ correspond to very different kinds of assumptions.  The theory consists of a fundamental quantum state and a prescription for the dynamics of the universe. These are the quantities that summarize the universal regularities exhibited by all physical systems within the universe. By contrast the xerographic distribution $\xi$ concerns a particular subsystem of the universe --- the HSI --- and its relation to other subsystems with the same data $D_0$. Priors representing independent assumptions about $T$ and $\xi$ are therefore natural; this implies a factorization
\begin{equation}
\label{factoredpriors}
P(T,\xi)=P_{\rm xd}(\xi)P_{\rm th}(T) .
\end{equation}
With these kind of priors we can compete different theories $T$ with the same typicality assumption $\xi$  and also compete different typicality assumptions $\xi$ with the same theory $T$.

There are those who are secure in the faith that the HSI must be typical of all other IGUSes in the universe despite the absence of any experimental evidence either for or against the supposition.  
They will chose a prior $P_{\rm xd}(\xi)$ that is unity for the $\xi$ that is uniform over the class of IGUSes, and zero elsewhere. 
Others, like the authors, who see $\xi$ as an assumption much like any other will allow different typicality assumptions to be competed against one another in the search for a successful  framework for the universe. The essence of science is to concede that there is at least some chance that any assumption is incorrect, and then check for that with experiment and observation.  The next section illustrates how this works in a simple model. 

\section{A Simple Cosmological Model} 
\label{RBmodel} 

The ideas in the preceding section can be illustrated by the simple red-blue model that we employed for illustrative purposes in \cite{HS07}.  Consider a model universe which consists of  $N$ cycles  in time, $k=1, \ldots, N$. In each cycle the universe may have one of two global properties: red $(R)$ or blue $(B)$, which could for example be thought of as two different possible values of the CMB temperature.  In each cycle, there is a probability $p_E$ for a physical system to exist $(E)$ that is able to observe this global property.  The model assumes that the observations are perfectly accurate, so that if red is observed in any cycle, then the universe is red in that cycle. It is further assumed that whether the universe is red or blue does not affect whether an observing system exists or not. 

Two competing theories of this model universe are proposed. One, {\it all red\/} ($AR$), 
in which all the cycles are red, and another, {\it some red\/} ($SR$), in which some number of particular cycles are red and the rest are blue.  
Suppose that we (the HSI) observe red. Our data $D_0$ is then $(E,R)$ and we seek to discriminate between the two theories on the basis of the likelihoods for this data. But as described above there are several choices for these likelihoods. 

The use of third-person likelihoods based on the theory $T$ alone was already discussed above and in \cite{HS07}.
All we know about this universe in third-person terms is that it exhibits at least one instance of a cycle with $(E,R)$ --- the one we are in. The third-person probability that there is at least one cycle with $(E,R)$ is the same as one minus the probability of the negation of this, which is probability that no observing system exists in a cycle in which the universe is red. Since the probability for an observing system {\it not\/} to exist in any one cycle is $1-p_E$, the likelihoods are
\begin{equation}
P(E,R|T) = 1 - (1-p_{E})^{N_R(T)}  
\label{rblikelihoods}
\end{equation}
where $N_R(T)$ is the number of red cycles in theory $T$, equal to $N$ when $T$ is  $AR$. 
Our data do not discriminate between the two theories when  $N_R$ is large enough in {\it both} theories to make $(1-p_E)^{N_R(T)}\ll 1$. Then
$P(E,R|T)\approx 1$ for both theories. 
Even though there may be  more many more red cycles in the $AR$ theory than the $SR$ theory, the probability that there is {\it at least one} red cycle with an observing subsystem approaches one for both theories when $N_R$ becomes large in both. The likelihoods are the same. Our little bit of data is not enough to discriminate between the two theories on the basis
of third-person probabilities.

The situation is different when we use first-person likelihoods to discriminate between different frameworks $(T,\xi)$. To illustrate this, consider a variety of typicality assumptions expressed as different xerographic distributions.   

 The simplest assumption we can make is that we are typical in the class of other instances of our data $D_0$. In the context of the present model that is the assumption that we are equally likely to be any of the instances of $(R,E)$ --- observers that exist in a red cycle. If there are $n_R$ such systems than the probability that we are the $A$th instance is $1/n_R$ and the corresponding xerographic distribution is
\begin{equation}
\xtypD=\frac{1}{n_R} \ , 
\label{xtypD}
\end{equation}
where the superscript `typD' means typical in the class with data $D_0$.  The first-person likelihoods for our data are then denoted by 
$P^\1p(E,R|T,\xtypD)$.  

In Appendix \ref{apprb}  we work through the transition from third to first-person probabilities using an explicit example of a typicality assumption and its associated xerographic distribution. That is instructive, but this model is so simple and symmetric that the results for the likelihoods follow from a few simple arguments.  

The xerographic distribution $\xtypD$ is non-zero only on instances of $D_0$ in red cycles. 
The probability that we see red, $P^\1p(R|E,T,\xtypD)$, is thus unity trivially and the probability that we see  blue is zero. But then, from the definition of conditional probabilities, 
\begin{equation} 
\label{conditionalD}
1= P^\1p(R|E,T,\xtypD)\equiv\frac{P^\1p(E,R|T,\xtypD)}{P^\1p(E|T,\xtypD)} \ . 
\end{equation}
The first-person probability that we exist  given that we are in a red cycle is the same as the third-person probability that at least one observing system (us) exists in a red cycle.   That is, from \eqref{rblikelihoods}
\begin{equation}
\label{exist}
P^\1p(E |T,\xtypD) = 1-(1-p_E)^{N_R(T)} .
\end{equation}
Thus the first-person likelihoods are given by 
\begin{equation}
\label{1plikeD}
P^\1p(E,R|T,\xtypD) = 1-(1-p_E)^{N_R(T)}. 
\end{equation}
The likelihoods for $R$ and $B$ correctly sum to unity.
These first-person likelihoods are unchanged from the third-person ones \eqref{rblikelihoods}, and no more able to  discriminate between theories than they were. 

We now turn our attention to other possible typicality assumptions represented by different xerographic distributions. First let's consider the  assumption that we are typical of {\it all} the observing systems that exist ($E$) in the model universe --- not just the ones that have our data $D_0$. Equivalently the assumption is that we are equally likely to be any of the observing systems that the universe exhibits. If there are $n_O$ observing systems, then the probability that we are the $A$th instance is $1/n_O$, and that defines the xerographic distribution 
\begin{equation}
\xtypO_A =\frac{1}{n_O} \ .
\label{xtypobs}
\end{equation} 
The first-person likelihoods for our data are then denoted by $P^\1p(E,R|\xtypO,T)$.
  
We are equally likely to exist in any cycle since they are all the same. The probability that we see red is therefore the probability that our cycle is red. This is $N_R(T)/N$.  Thus 
\begin{subequations}
\label{1p}
\begin{equation}
\label{1pR}
P^\1p(R|E,\xtypO,T) = N_R(T)/N   .
\end{equation}
Similarly the probability that we see blue is
\begin{equation}
\label{1pB}
P^\1p(B|E,\xtypO,T) = N_B(T)/N   .
\end{equation} 
\end{subequations}
We can now proceed as we did above from \eqref{conditionalD} to \eqref{1plikeD}. The result for the first-person likelihoods is 
\begin{equation}
\label{1plike}
P^\1p(E,R|\xtypO,T) = \frac{N_R(T)}{N}[1-(1-p_E)^N]. 
\end{equation}
The ratio of the likelihoods for the two theories is 
\begin{equation}
\label{compTtyp}
\frac{P^\1p(E,R|\xtypO,AR)}{P^\1p(E,R|\xtypO,SR)} = \frac{N}{N_R(SR)} >1
\end{equation}
Thus, assuming equal priors for the two theories, $AR$ is always favored even if $N_R(SR)$ becomes arbitrarily large provided there are at least some blue cycles. 

In the above examples the two theories $AR$ and $SR$ are competed with the same typicality assumption. But it is also possible to compete different typicality assumptions for the same theory. 
Suppose we assign unit prior probability to the theory $SR$ and equal priors to $\xtypO$ and $\xtypD$. 
From \eqref{1plike} and  \eqref{1plikeD} we find 
\begin{equation}
\label{compxi}
\frac{P^\1p(E,R|\xtypO,SR)}{P^\1p(E,R|\xtypD,SR)} = \frac{N_R(SR)}{N}\frac{1-(1-p_E)^N}{1-(1-p_E)^{N_R(SR)}} <1.
\end{equation}
Thus an assumption of typicality in the class of our data $D_0$  does a better job of explaining our data (trivially). 

Certain theoretical models may imply that the number of instances of our data $N$ is infinite. An example is provided by the infinite number of Hubble volumes  on a surface of homogeneity inside a Coleman-De Luccia bubble of false vacuum \cite{CdL80}. The results above are well defined provided that the fractions $f_R(T)$ and $f_B(T)$ of red and blue Hubble volumes are well defined in the competing theories $T$.  For instance \eqref{1p} becomes
\begin{equation}
\label{1pf}
P^\1p(R|E,\xtypO,T) = f_R(T),  \quad  
P^\1p(B|E,\xtypO,T) = f_B(T) .    
\end{equation} 
and similarly with the other formulae. Indeed, the expressions are generally simpler than for the finite case since the probability that there is at least one instance of our data \eqref{exist} becomes exactly unity. 

This set of examples shows that frameworks $(T, \xi)$ are an umbrella formalism for organizing, comparing, and (most importantly) making explicit a number of different assumptions that are commonly made in the quantum cosmology of a very large universe. Some of these possible
assumptions will be discussed in more detail in Section \ref{measures} and Appendix \ref{pageapp}. 

\section{Predictivity}
\label{pred}

First let us consider theories $T$ where there is at most one copy of our data $D_0$.  No xerographic distribution is needed.  In this setting, a theory is {\it predictive} if the likelihood is high for some piece of data that we might acquire in the future, given the theory and the data that we have acquired in the past. To make this more concrete, consider a simplified situation in which a stream of data $d_{-n},\ldots,d_{-1},d_0$ is acquired at a sequence of times $t_{-n},\ldots,t_{-1},t_0$, where $t_0$ denotes the present time, defined relative to the local clocks provided by each particular occurrence of $D_0$. 
Assuming that all this data is accessible now, our present data is  the union, $D_0=d_{-n}\cup\ldots\cup d_0$.  Let $P^\1p(q_1|T,D_0)$ be the first-person  likelihood that at some future time $t_1$ we will acquire some piece of data $q_1$; $q_1$ is a subset of all the data $d_1$ that we acquire at that time.  A theory  is {\it predictive} if, for {\it some} kinds of data $q_1$, the likelihood $P^\1p(q_1| T,D_0)$ 
is sharply peaked around particular values $\bar q_1$.   

There is obvious motivation to favor theories that are predictive and provide a coherent story connecting past data that we currently  have to future data that we may acquire. The utility of physical theory lies in its predictive power. Theories that are predictive are also testable in the sense that they are falsifiable when new data disagrees with that predicted. Indeed, the whole history of science can be read as the search for predictive theories. 

We turn now to theories of large universes with replication of our data $D_0$.  As already noted, in this case we can only make predictions if we specify a xerographic distribution $\xi$ in addition to a conventional theory $T$.  Once the xerographic distribution is specified, we can (at least in principle) compute the likelihood $P^\1p(q_1|T,\xi,D_0)$ that we will acquire a piece of data $q_1$ at a future time $t_1$.   A framework $(T,\xi)$ is {\it predictive\/} if, for {\it some} kinds of data $q_1$, the likelihood $P^\1p(q_1|T,\xi,D_0)$ is sharply peaked around particular values $\bar q_1$.  

To compute $P^\1p(q_1|T,\xi,D_0)$, we first compute the third-person likelihood that the data subset $q_1$ is found at time $t_1$ at the $A$th location of the data $D_0$; for this we need only the theory $T$ and the data $D_0$.  Denote this likelihood by $P(q_1 @ A|T,D_0@A)$,
where $x@A$ means that data $x$ occurs at location $A$.
Then the likelihood that `we' obtain $q_1$ is 
\begin{equation}
P^\1p(q_1|T,\xi,D_0)=\sum_A \xi_A P(q_1 @ A|\xi, T,D_0 @ A).
\label{Pxi} 
\end{equation}
Without an assumed xerographic distribution, no prediction whatsoever can be made about what `we' will see in the future.  A physical theory that is considered `complete' in the usual sense
(such as a specified quantum state and a rule for its evolution, or even a fully deterministic classical theory plus initial data) is insufficient to determine the xerographic distribution, which must therefore be chosen as an additional ingredient of the theoretical framework. 

\section{Making Theories Predictive with Typicality Assumptions}

Frameworks $(T,\xi)$ with the same theory $T$ but different assumptions for the xerographic distribution $\xi$ can be compared and tested.   Scientists favor frameworks that are predictive, that is, that generate a stream of future predictions and hence are testable.

In this section we give two examples how the predictivity of frameworks $(T,\xi)$ with a fixed theory $T$
can be affected by different choices of $\xi$ that reflect different assumptions about the typicality of our data $D_0$ in two different situations. 

\subsection{Boltzmann Brains}

Thermal or vacuum fluctuations could replicate our data \cite{BBrefs,Page06}. In one simplified model, a spatially closed universe originates in a Big Bang and eventually enters a de Sitter phase that persists forever. It is assumed that there is a very tiny probability per unit spacetime volume that a `brain' could fluctuate into existence.
Such fluctuations are called `Boltzmann brains' (BBs) or `freak observers' \cite{BBrefs}.  
Since the spacetime volume is infinite in this model (even though the spatial volume is finite), it is assumed that an infinite number of BBs will be fluctuated into existence. A tiny fraction of this infinite number have the same data $D_0$ that we do. How do we know that we are not one of them?  This is the Boltzmann brain problem. 

Let us accept this scenario uncritically, and ask only if there is an assumption for the xerographic distribution $\xi$ for which the framework is predictive. 

An assumption that we are typical, $\xi_A=1/N$, does not result in a predictive framework. In that case, we are much more likely to be a BB than an ordinary observer (OO). The BBs are deluded (e.g \cite{Got08}); their data suggest that they are $13.7\,$Gyr from a Big Bang, but typically they are much further away. In contrast to ordinary observers that have $13.7\,$Gyr of history, the subsequent observations of any particular BB is overwhelmingly likely to be disordered, and inconsistent with its apparent history.  Future data is thus uncorrelated with $D_0$, and firm predictions cannot be made.

Now let us consider a nonuniform $\xi$; that is, we assume that we are atypical. In particular, we suppose that $\xi_A$ is nonzero only for locations sufficiently close to a Big Bang to make it much more likely that we are OOs rather than BBs. In this case, the framework is predictive in the usual way. 

The answer to the question `How do we know that we are not BBs?' is this: we do not know. But if we assume a xerographic distribution $\xi$ such that we are not likely to be BBs, then we get a predictive, testable framework  $(T,\xi)$. Confirmation of the predictions of this framework by a series of observations then support the original assumption of atypicality.\footnote{Others have expressed related ideas, e.g \cite{Got08}.}
 
\subsection{Laboratory Experiment} 

Consider a laboratory experiment to determine the value of a fundamental constant $\alpha$ by a sequence of measurements.  We consider a class of theories that predict that $\alpha$ is constant throughout spacetime, but predict different values for it.  In a very large universe, the experiment will be replicated in many different locations $x_A$ in all essential details. For the vast majority of these experiments the mean value $\bar\alpha$ of the sequence of outcomes will be close to the true  value of the constant to the accuracy of the experiment. But for a tiny fraction of these experiments, the sequence of measurement outcomes are mistaken --- their mean value $\bar\alpha$ is far from the true value --- just by the chances of statistics. How do we know that our particular sequence of results is not one of these? 

The answer lies in the evidence from further measurements. The probabilities for further measurements can be predicted from the previous ones once a xerographic distribution $\xi$ is specified. An assumption that that our experiment is typical of all the others results in a theoretical framework $(T(\bar\alpha), \xi_A{=}1/N)$, where $T(\bar\alpha)$ is the theory that predicts that the value of the constant is the mean value of the previous outcomes, $\bar\alpha$. The prediction of this framework is that the next measurement should yield $\bar\alpha$ to within statistical error.  Confirmation of this prediction by further measurements supports the framework $(T(\bar\alpha), \xi_A{=}1/N)$, both as to the value of $\alpha$ and the assumption of typicality.  

In the way the problem has been set up, there is no evident variable with which to make an assumption of atypicality. The physical situations of all the experiments have been assumed to be the same, in contrast to the different situations represented by BBs and OOs.  As emphasized correctly be a number of authors (e.g.~\cite{GV07}), some kind of typicality is assumed implicitly in the analysis of every laboratory experiment. Here we have made that assumption explicit in terms of the xerographic distribution. 

\section{Measures for Cosmology}
\label{measures}

In models of eternally inflating cosmologies, relative probabilities for different kinds of physical situations are defined in terms of the ratios of the number of times they occur. Examples are the occurrence of different kinds of `bubble universes'  and the ratio of the number of BBs to OOs. Since the numbers are typically infinite, a `measure' is required to define these fractions; 
without a measure, third-person probabilities are ill-defined.  

Ideally, such a measure would emerge unambiguously from an underlying theory (such as string theory). In this case, the `measure problem' would be solved, and third-person probabilities could be computed.

However, a xerographic distribution would still be needed in order to define and compute first-person probabilities.  Solving the measure problem does not remove the need to choose a xerographic distribution, but does make the choice of a uniform xerographic distribution well defined.  

In some models (e.g. \cite{HHH}), there is a natural choice for the measure.  However, in more general contexts there is as yet no consensus for how to determine the measure from the underlying theory, or even whether this is possible (see e.g. \cite{GV07,measrefs} for discussions).  If it turns out that the measure is not computable from the underlying theory, then whatever freedom remains in the choice of the measure can and should be incorporated into the choice of the xerographic distribution.   

Starting from a different perspective, Page \cite{Page08ab} has also argued that the quantum wave function alone contains insufficient information for the calculation of probabilities of subsequent observations by a particular observer when there are multiple copies of that observer. His work is discussed in Appendix \ref{pageapp}.

\section{Conclusion}
 
The possibility that our data may be replicated exactly elsewhere in a very large universe profoundly affects the way science must be done. 
 
Central to cosmology are the third-person probabilities for properties of the universe given a theory $T$ of its quantum dynamics and quantum state. But of even greater interest are the first-person probabilities for the results of observation carried out by us --- a particular subsystem of the universe --- conditioned on our existing data.  These probabilities are the means to test any prescription for prediction.  In a very large universe, where our data will be replicated elsewhere with significant probability, predicting these probabilities requires not only a theory $T$ but also but also  a probability distribution on the set of copies of us. 
This xerographic distribution cannot, even in principle, be determined from a theory of the dynamics and the quantum state
since such a theory has nothing to say about which copy is `us'.  It is only the theoretical framework consisting of both a theory and an assumed xerographic distribution that is predictive and testable by observation.

In cosmology we should favor theoretical frameworks that generate a stream of predictions from our data that are confirmed by subsequent observation. The authors believe that choices of both the theory $T$ and the xerographic distribution $\xi$ should be competed against other alternatives, and that no element of the theoretical framework should be assigned a unit probability.  

Ideas that imply particular notions of typicality (such as the `Copernican principle', `anthropic principle', or the `principle of mediocrity') cannot be universal laws of nature if only because they refer to a negligibly minor subsystem of the universe: `us'. In the present context, these are simply notions that can motivate a particular choice of the xerographic distribution.  

It is no surprise that information about us is required to make predictions for our observations. Our data suggest that we are located some $13.7\,$Gyr from a Big Bang. To make a reliable prediction from that information, we have to assume that it truly describes our physical situation. 

If the universe is rife with delusion, we must assume that we are atypical in order to have predictive and testable scientific theories. Indeed, it is only by making such assumptions that we are able to do science  in a very large universe. We imagine that even Copernicus would have  agreed that it was necessary to assume that Ptolemy was not deluded in his observations of the planets.

\acknowledgments
We thank Raphael Bousso, Brandon Carter, Sean Carroll, Ben Freivogel, Steve Giddings, Alan Guth, Thomas Hertog, Matthew Kleban, Don Page, Steve Shenker, and Alex Vilenkin for numerous helpful conversations.
This work was supported in part by the National Science Foundation under grants PHY05-55669
and PHY07-57035.

\appendix

\section{The Work of Page as an Example}
\label{pageapp}

As mentioned above, the schema for quantum cosmological prediction developed here provides a common framework for the discussion of different prescriptions for  science in a very large universe. Prescriptions for which notions of first- and third-person probabilities can be distinguished, and a xerographic distribution identified. 

Notable among these different presciptions are those found in the extensive contributions of Don Page \cite{Page08ab,Page09a,Page09b}. In this Appendix we attempt to fit at least some parts of his ideas into the present context, in part to address his criticisms of our earlier 
work \cite{HS07}. 

Page reaches the conclusion that Born's rule of usual quantum theory must be modified to apply to a very large universe where our data have a significant chance of being replicated elsewhere. We disagree with this conclusion as we now explain.

We begin with Page's discussion of Born's rule. In \cite{Page09b}, he says that ``a goal of science is to produce complete theories $T_i$ that each predict normalized probabilities $P_j(i)$ for observations $O_j$''. These are the probabilities for observations made by ``the observer''.  We interpret these as first-person probabilities for alternative outcomes of our observations. For example they might be the probabilities for alternative CMB temperature maps or the alternatives red and blue in the R/B model. The sum of the probabilities  of an exhaustive set of exclusive outcomes is of course $1$.   This is illustrated explicitly by \eqref{1p}. 

We agree with Page that first-person probabilities are not specified by the quantum state of the universe alone. As we have argued, a xerographic distribution is needed to to connect them to the third-person probabilities that are specified by the quantum state. However this does not mean that `Born's rule dies.' Born's rule is alive, well, and essential for constructing third-person probabilities. 

Page suggests various candidates for replacing Born's rule.  In his nomenclature \cite{Page09a}, our xerographic distribution for the assumption that we are typical of all the observing systems that exist appears to be equivalent to Page's rule 5, which he calls 
`observational averaging'.

Constructions such as those of Page or the theoretical frameworks discussed in this paper are tested by comparing observations with the first-person predictions for them. The third-person probabilities for features of the universe are essential for computing these. For instance, both Page's candidates for replacing Born's rule and the examples of xerographic distributions in this paper rely on an assumption of classical spacetime to make notions of location meaningful. But classical spacetime is neither fundamental nor inevitable in a quantum theory of gravity. Whether a quantum state predicts an ensemble of alternative classical spacetimes is a question of whether the third-person probabilities are high for correlations in time governed by the Einstein equation for suitably coarse grained histories of geometry and matter fields (e.g. \cite{HHH}). Third-person probabilities are  thus not dispensable; they are essential for the understanding of our quantum universe. 

\section{Third to First in the R/B Model with a Xerographic Distribution}
\label{apprb}

In this appendix the first-person likelihoods \eqref{1plike} in the red-blue model of Section  \ref{RBmodel} are derived by explicitly considering the form of the xerographic distribution $\xtypO$ without invoking the symmetry of the model directly.

We start with the third-person probability $P(n_O,n_R|T)$ that of the $N$ total cycles, $n_O$ are occupied by observing systems, with $n_R$ of these in red cycles. The probability that there are $n_O$ cycles occupied by observing systems is $p_E^{n_O}(1-p_E)^{N-n_O}$ multiplied by the number of ways of arranging $n_R$ observing systems in $N_R$ red cycles and the number of ways of arranging $n_B=n_O-n_R$ observing systems in $N_B=N-N_R$ blue cycles. This is
\begin{equation}
\label{3pnnR}
P(n_O,n_R|T) = {N_R \choose n_R}{N-N_R \choose n-n_R} \pe^{n_O} (1-\pe)^{N-n_O}
\end{equation}
The theoretical framework consists of the theory $T$ and a xerographic distribution that assumes we are a typical in the class of observing systems. The $n$ observing systems can be labeled by an index $A$ that runs from $1$ to $n$. The xerographic distribution that says we are equally likely to be any one of these systems is
\begin{equation}
\xtypO_A = \frac{1}{n_O} ,
\label{xtypO}
\end{equation}
which satisfies
\begin{equation}
\sum_{A=1}^{n_O} \xtypO_A  = 1 .
\label{xdistsum}
\end{equation}
With these assumptions the first-person likelihood $P^\1p(E,R|\xi^{\rm typO},T)$ that we exist and observe red is $\xtypO_A$ times the third-person probability $P(n_O,n_R|T)$ summed over $A$ and the alternative configurations specified by $(n_O,n_R)$.  In general the third-person probabilities would depend on which cycles are occupied and hence on $A$. But in the present case where they are all identical there is no such dependence. The result is
\begin{equation}
\label{1plikex}
P^\1p(E,R|\xtypO,T) = \sum_{n_O=1}^N \sum_{n_R=1}^{n_O} \sum_{A=1}^{n_R}
 \xtypO_A P(n_O,n_R|T)
\end{equation}
Substituting in \eqref{3pnnR} and \eqref{xtypO} and performing the sum over $A$, we get
\begin{align}
\label{1plikexO}
P^\1p(E,R|\xtypO,T) &= \sum_{n_O=1}^N \sum_{n_R=1}^{n_O} \frac{n_R}{n_O}
{N_R \choose n_R}{N-N_R \choose n_O-n_R} \pe^{n_O}(1-\pe)^{N-n_O}
\nonumber \\ 
&=\sum_{n_O=1}^N \frac{N_R}{n_O} \sum_{n_R=1}^{n_O} {N_R-1 \choose n_R -1}{N-N_R \choose n_O-n_R} \pe^{n_O}(1-\pe)^{N-{n_O}}
\nonumber \\ 
&=\sum_{n_O=1}^N \frac{N_R}{n_O} {N-1 \choose n_O-1} \pe^{n_O}(1-\pe)^{N-n_O}
\nonumber \\ 
&=\frac{N_R}{N}  \sum_{n_O=1}^N  {N\choose n_O}  \pe^{n_O}(1-\pe)^{N-n_O}  
\nonumber \\
&=\frac{N_R}{N} [1-(1-p_E)^N] \ ,
\end{align}
where the third line follows from Vandermonde's identity. 

We see that \eqref{1plikexO} is the same as \eqref{1plike}. The argument based on symmetry given in Section \ref{RBmodel} is evidently a more efficient way of getting this result. A general case without symmetry would be even more complicated. Suppose for example there was a different value of $p_E$ for each cycle. Then the third-person probabilities would depend on which cycles were occupied and not just on the total number of them as here. The prescription for computation however would be essentially the same: calculate the third-person probabilities for a configuration of occupied and unoccupied cycles; multiply by the xerographic distribution; sum over the possible configurations.

\end{document}